\documentclass[10pt]{article}
\usepackage{graphicx}
\usepackage{relsize}

\newcommand{\remove}[1]{}

\newcommand{\op}[1]{{\textsf{\textit{\smaller{#1}}}}}

\usepackage{color}
\usepackage{url}

\begin{document}
\begin{titlepage}

\title{Lock-free Concurrent Data Structures\footnote{To appear in ``Programming Multi-core and Many-core Computing Systems'', eds. S. Pllana and F. Xhafa, Wiley Series on Parallel and Distributed Computing.}}
\newcommand{\affilmark}[1]{\rlap{\textsuperscript{\itshape#1}}}

\author{Daniel Cederman\affilmark{1} \and Anders Gidenstam\affilmark{2} \and
Phuong Ha\affilmark{3} \and H{\aa}kan Sundell\affilmark{2} \and Marina Papatriantafilou\affilmark{1} 
\and Philippas Tsigas\affilmark{1}
\\
\\
\affilmark{1} Chalmers University of Technology, Sweden\\
\affilmark{2} University of Bor{\aa}s, Sweden\\
\affilmark{3} University of Troms\o, Norway}
\date{}

\maketitle \thispagestyle{empty}
\end{titlepage}

\section{Introduction}
\index{data structures}
\index{concurrent data structures}
\index{concurrency}
Concurrent data structures are the data sharing side of parallel programming.
Data structures give the means to the program to
store data but also provide operations to the program to access and manipulate these data. These operations are
implemented through algorithms that have to be efficient. In the sequential setting, data structures are crucially important for the performance of the respective computation. \index{parallel programming}
In the parallel programming setting, their importance becomes more crucial because of the increased use of data and resource sharing for utilizing parallelism. In parallel programming, computations are split into subtasks in order to introduce parallelization at the control/computation level. To utilize this opportunity of concurrency, subtasks share data and
various resources (dictionaries, buffers, and so forth).
This makes it possible for logically independent programs to share various resources and data structures. A subtask that wants to update a data structure, say add an element into a dictionary, that operation may be logically independent of other subtasks that use  the same dictionary.

\index{mutual exclusion}
Concurrent data structure designers are striving to maintain consistency of data structures while keeping the use of mutual exclusion and expensive synchronization to a minimum, in order to prevent the data structure from becoming a sequential bottleneck. Maintaining consistency
in the presence of many simultaneous updates is a complex task.
Standard implementations of data structures are based on locks
in order to avoid inconsistency of the shared data due to concurrent modifications. In simple terms, a single lock around the whole data structure may create a bottleneck in the program where all of the tasks serialize, resulting in a loss of parallelism because too few data locations are concurrently in use.
\index{deadlock} \index{priority inversion} \index{convoying} \index{lock}
Deadlocks, priority inversion, and convoying are also side-effects of locking.
The risk for deadlocks
makes it hard to compose different blocking data structures since it is not always
possible to know how closed source libraries do their locking. 
It is worth noting that in graphics processors (GPUs) locks are not recommended for designing concurrent data structures.
GPUs prior to the NVIDIA Fermi architecture do not have writable caches, so for
those GPUs, repeated checks to see if a lock is available or not require expensive repeated
accesses to the GPU's main memory. While Fermi GPUs do support writable caches, there is no guarantee that the thread scheduler
will be fair, which can make it difficult to write deadlock-free locking code. OpenCL \index{OpenCL}
explicitly disallows locks for these and other reasons.

\index{lock-free}
Lock-free implementations of data structures support concurrent access.
They do not involve mutual exclusion and make sure that all steps
of the supported operations can be executed concurrently.
\index{optimistic synchronization}
Lock-free implementations employ an optimistic conflict control
approach, allowing several processes to access the shared data object at the
same time. They suffer delays only when
there is an actual
conflict between operations that causes
some operations to retry.
This feature allows lock-free algorithms to scale much better when
the number of processes increases.

An implementation of a data structure is called {\em lock-free} if it allows multiple processes/threads to access the data structure concurrently and also guarantees that
at least one operation among those
finishes in a finite number of its own steps regardless of the state of the other operations.
\index{linearizability}
A consistency (safety) requirement for
lock-free data structures is {\em linearizability} \cite{HerW90},
which ensures that each operation
on the data appears to take effect instantaneously during its actual duration and
the effect of all operations are consistent with the
object's
 sequential specification.
Lock-free data structures offer several advantages over their
blocking counterparts, such as being immune to deadlocks,
priority inversion, and convoying, and have been shown to
work well in practice in many different settings \cite{TsiZ02,SunT02}. They have been included \index{Threading Building Blocks}
in Intel's Threading Building Blocks Framework \cite{tbb}, \index{NOBLE}
the NOBLE library \cite{SunT02} and the Java concurrency package  \cite{javacp},
and will be included in the forthcoming parallel extensions
to the Microsoft .NET Framework \cite{msconc}. \index{.NET} \index{C++} \index{Java}
They have also been of interest to designers of languages such as C++ \cite{DecPS06} and Java~\cite{javacp}.

This chapter has two goals.
The first and main goal is to provide a sufficient background and intuition to help the interested reader to navigate in the complex research area of lock-free data structures. The second goal is to offer the programmer familiarity to the subject that will allow her to use truly concurrent methods.

The chapter is structured as follows.
First we discuss the fundamental and commonly-supported synchronization primitives on which efficient lock-free data structures rely.
Then we give an overview of the research results on lock-free data structures that appeared in the literature with a short summary for each of them.
The problem of managing dynamically-allocated memory in lock-free concurrent data structures and general concurrent environments is
discussed separately. Following this is a discussion on the idiosyncratic architectural features of
graphics processors that are important to consider when designing
efficient lock-free concurrent data structures for this emerging area. 

\section{Synchronization Primitives}\label{SecSyncPrim}
\index{synchronization primitives}
To synchronize processes efficiently, multi-/many-core systems usually support certain synchronization primitives.
This section discusses the fundamental synchronization primitives, which typically read the value of a {\em single} memory word, modify the value and write the new value back to the word {\em atomically}.

\subsection{Fundamental synchronization primitives}

The definitions of the primitives are described in Figure~\ref{fig:intro_primitives}, where
$x$ is a memory word,
$v, old, new$ are values and $op$ can be operators $add$, $sub$, $or$, $and$ and $xor$. Operations between angle brackets $\langle\rangle$ are executed atomically.
\begin{figure}[tbh]
\hrule
\centering
\small
\begin{tabular}{p{2in}p{1in}}
\begin{tabbing}
{\bf TAS}\=$(x)$ {\em /* test-and-set, init: $x \leftarrow 0$ */}\\
\> $\langle oldx \leftarrow x$; $x \leftarrow 1$; {\bf return} $oldx; \rangle$ \\

\\

{\bf FAO}\=$(x,v)$ {\em /* fetch-and-op */} \\
\> $\langle oldx \leftarrow x$; $x \leftarrow op(x,v)$; {\bf return} $oldx; \rangle$\\

\\

{\bf CAS}\=$( x, old, new)$ {\em /* compare-and-swap */} \\
\> $\langle$ \= {\bf if}$(x=old)$ $\{ x \leftarrow new$; {\bf return}$(true);\}$\\
\>\>{\bf else return}$(false)$; $\rangle$

\end{tabbing}
&
\begin{tabbing}
{\bf LL}\=$(x)$ {\em /* load-linked */}\\
\> $\langle return~the~value~of~x~so~that$\\
\> $it~may~be~subsequently~used$\\
\> $with$ \bf{SC} $\rangle$\\

{\bf SC}$(x,v)$ {\em /* store-conditional */}\\
\> $\langle$ \= {\bf if} $(no~process~has~written~to~x$ \\
\> $since~the~last$ {\bf LL}$(x))$ $\{ x \leftarrow v$; \\
\>\> {\bf return}$(true) \}$;\\
\>\> {\bf else return}$(false)$; $\rangle$
\end{tabbing}
\end{tabular}
\hrule
\caption{Synchronization primitives}\label{fig:intro_primitives}
\end{figure}

Note that there is a problem called the \index{ABA problem} {\em ABA problem} that may occur with the {\em CAS} primitive. The reason is that the {\em CAS} operation can not detect if a variable was read to be A and then later changed to B and then back to A by some concurrent processes. The {\em CAS} primitive will perform the update even though this might not be intended by the algorithm's designer. The {\em LL/SC} primitives can instead detect any concurrent update on the variable between the time interval of a {\em LL/SC} pair, independent of the value of the update.

\subsection{Synchronization power}
The primitives are classified according to their synchronization power or \index{consensus number} {\em consensus number}~\cite{Her91}, which is, roughly speaking, the maximum number of processes for which the primitives can be used to solve a {\em consensus problem} in a fault tolerant manner. In the consensus problem, a set of $n$ asynchronous processes, each with a given input, communicate to achieve an agreement on one of the inputs. A primitive with a consensus number $n$ can achieve consensus among $n$ processes even if up to $n-1$ processes stop~\cite{Turek92}.

\index{synchronization primitives!compare-and-swap (CAS)}
 \index{synchronization primitives!load-linked/store-conditional (LL/SC)} 
 \index{synchronization primitives!fetch-and-op (FAO)}
  \index{synchronization primitives!test-and-set (TAS)}
According to the consensus classification, read/write registers have consensus number 1, i.e. they cannot tolerate any faulty processes in the consensus setting. \index{compare-and-swap (CAS)} \index{load-linked/store-conditional (LL/SC)} \index{fetch-and-op (FAO)} \index{test-and-set (TAS)}
There are some primitives with consensus number 2 (e.g. {\em test-and-set (TAS)} and {\em fetch-and-op (FAO)}) and some with infinite consensus number (e.g. {\em compare-and-swap (CAS)} and {\em load-linked/store-conditional (LL/SC)}).
It has been proven that a primitive with consensus number $n$ cannot implement a primitive with a higher consensus number in a system of more than $n$ processes \cite{Her91}. For example, the {\em test-and-set} primitive, whose consensus number is two, cannot implement the {\em compare-and-swap} primitive, whose consensus number is unbounded, in a system of more than two processes.

\subsection{Scalability and Combinability}
\index{scalability}
As many-core architectures with thousands of cores are expected to be our future chip architectures \cite{Asa06}, synchronization primitives that can support scalable thread synchronization for such large-scale architectures are desired.
In addition to synchronization power criterion, synchronization primitives can be classified by their scalability or {\em combinability} \cite{KruRS88}.
Primitives are combinable if their memory requests to the same memory location (arriving at a switch of the processor-to-memory interconnection network) can be combined into {\em only one} memory request. Separate replies to the original requests are later created from the reply to the combined request (at the switch). The combining technique has been implemented in the NYU Ultracomputer \cite{GotGKMRS82} and the IBM RP3 machine \cite{PfiBGHKMMNW85}, and has been shown to be a scalable technique for large-scale multiprocessors to alleviate the performance degradation due to a synchronization ``hot spot''.
The set of combinable primitives includes {\em test-and-set}, {\em fetch-and-op} (where $op$ is an associative operation or boolean operation), blocking full-empty bits \cite{KruRS88} and non-blocking full-empty bits \cite{HaTA09_OPODIS}.
For example, two consecutive requests {\em fetch-and-add}$(x,a)$ and {\em fetch-and-add}$(x,b)$ can be combined into a {\em single} request {\em fetch-and-add}$(x,a+b)$. When receiving a reply $oldx$ to the combined request {\em fetch-and-add}$(x,a+b)$, the switch at which the requests were combined, creates a reply $oldx$ to the first request {\em fetch-and-add}$(x,a)$ and a reply $(oldx + a)$ to the successive request {\em fetch-and-add}$(x,b)$.

The $CAS$ primitives are not combinable since the success of a $CAS(x,a,b)$ primitive depends on the current value of the memory location $x$. For $m$-bit locations (e.g. 64-bit words), there are $2^m$ possible values and therefore, a combined request that represents $k$ $CAS(x,a,b)$ requests, $k < 2^m$, must carry as many as $k$ different checking-values $a$ and $k$ new values $b$.
The $LL/SC$ primitives are not combinable either since the success of a $SC$ primitive depends on the state of its reservation bit at the memory location that has been set previously by the corresponding $LL$ primitive. Therefore, a combined request that represents $k$ $SC$ requests (from different processes/processors) must carry as many as $k$ store values.

\subsection{Multi-word Primitives}
Although the {\em single-word} hardware primitives are conceptually powerful enough to support higher-level synchronization, from the programmer's point of view they are not as convenient as {\em multi-word} primitives. The multi-word primitives can be built in hardware \cite{KelF85, DieH_2008, ChaCEKLYZT_Micro09}, or in software (in a lock-free manner) using single-word hardware primitives \cite{AndM95, FraH_TOCS07, HaT04, IsrR94, Moi97b, ShaT95}.
\index{transactional memory}
Sun's third generation chip-multithreaded (CMT) processor called Rock is the first processor supporting transactional memory in hardware \cite{ChaCEKLYZT_Micro09}. The transactional memory is supported by two new instructions {\em checkpoint} and {\em commit}, in which {\em checkpoint} denotes the beginning of a transaction and {\em commit} denotes the end of the transaction. If the transaction succeeds, the memory accesses within the transaction take effect atomically. If the transaction fails, the memory accesses have no effect.

\index{synchronization primitives!NB-FEB}
\index{synchronization primitives!Advanced Synchronization Facility (ASF)}
\index{Advanced Synchronization Facility (ASF)}
Another emerging construct is the Advanced Synchronization Facility (ASF), an experimental AMD64 extension that AMD's Operating System Research Center develops to support lock-free data structures and software transactional memory \cite{DieH_2008}. ASF is a simplified hardware transactional memory in which all memory objects to be protected should be statically specified before transaction execution. Processors can protect and speculatively modify up to 8 memory objects of cache-line size. \index{NB-FEB}
There is also research on new primitives aiming at identifying new efficient
and powerful primitives, with the non-blocking full/empty bit (NB-FEB)
being an example that was shown to be as powerful as $CAS$
or $LL/SC$ \cite{HaTA09_OPODIS}.

\section{Lock-Free Data Structures}

The main characterization on which one can classify the various implementations of lock-free data structures available in the literature, is what \textit{abstract data type} \index{abstract data type} that it intends to implement. For each abstract data type there are usually numerous implementations, each motivated by some specific targeted purposes, where each implementation is characterized by the various properties that it fulfills to different amounts. As many of these properties are orthogonal, for each specific implementation, one or more properties are often strengthened at the cost of some others. Some of the most important properties that differentiate the various lock-free data structure implementations in the literature are:

\noindent{\bf Semantic fulfillments} Due to the complexity of designing lock-free data structures it might not be possible to support all operations normally associated with a certain abstract data type. Hence, some algorithms omit a subset of the normally required operations and/or support operations with a modified semantics.

\noindent{\bf Time complexity} Whether an operation can terminate in a time (without considering concurrency) that is linearly or logarithmically related to e.g. the size of the data structure, can have significant impact on performance. Moreover, whether the maximum execution time can be determined at all or if it can be expected in relation to the number of concurrent threads is of significant importance to time-critical systems (e.g. real-time systems).

\index{scalability}
\noindent{\bf Scalability} Scalability means showing some performance gain with increasing number of threads. Synchronization primitives are normally not scalable in themselves; therefore it is important to avoid unnecessary synchronization. Israeli and Rappoport \cite{IsrR94} have defined the term 
\index{disjoint-access-parallelism}
\textit{disjoint-access-parallelism} to identify algorithms that do not synchronize on data that is not logically involved simultaneously in two or more concurrent operations.

\noindent{\bf Dynamic capacity} In situations where it can be difficult to determine the maximum number of items that will be stored in a data structure, it is necessary that the data structure can dynamically allocate more memory when the current capacity is about to be exceeded. If the data structure is based on statically allocated storage, capacity is fixed throughout the lifetime of the data structure.

\noindent{\bf Space complexity}
\remove{Whether the memory required for representing data is linear or exponential related to the actual amount of data stored can be significant.}
Some algorithms can guarantee an upper bound of memory required, while some others can transiently need an indefinite amount depending on the concurrent operations' invocation order, and can thus not be deterministically determined.

\noindent{\bf Concurrency limitations} Due to the limitations (e.g. consensus number) of the chosen synchronization primitives, some or all operations might not allow more than a certain number of concurrent invocations.

\index{synchronization primitives!DCAS}
\index{synchronization primitives!CASN}
\index{synchronization primitives!MWCAS}
\noindent{\bf Synchronization primitives} Contemporary multi-core and many-core systems typically only support single-word \op{CAS} or weak and non-nestable variants of \op{LL/SC} (cf. section \ref{SecSyncPrim}). However, many algorithms for lock-free data structure depend on more advanced primitives as e.g. double-word \op{CAS} (called e.g. \index{DCAS} \op{DCAS} or \op{CAS2}), ideal \op{LL/SC} or even more complex primitives. These algorithms then need (at least) one additional abstraction layer for actual implementation, where these more advanced primitives are implemented in software using another specific algorithm. The \op{LL/SC} primitives can be implemented e.g. by \op{CAS} \cite{IsrR94, AndM95,Moi97,JayP03,Mic04c}. Multi-word \op{CAS} (called e.g. \index{CASN} \index{MWCAS} \op{MWCAS} or \op{CASN}) can be implemented e.g. by \op{CAS} \cite{HarFP02,Sun09} or by \op{LL/SC} \cite{IsrR94,AndM95,ShaT95,Moi97b,HaT04}.

\noindent{\bf Reliability} Some algorithms try to avoid the ABA problem by the means of e.g. version counters.
As these counters are bounded and can overflow, there is a potential risk of the algorithm to actually perform
 incorrectly and possibly cause inconsistencies.
 Normally, by design this risk can be kept low enough that it fits for practical purposes, although the risk increases as the computational speed increases. Often, version counters can be removed altogether by the means of proper memory management.

\noindent{\bf Compatibility and Dependencies} Some algorithms only work together with certain memory allocators and reclamation schemes, specific types (e.g. real-time) of system-level process scheduler,
or require software layers or semantic constructions only found in certain programming languages (e.g. Java).

\subsection{Overview}
The following sections include a systematic overview of the research result in the literature.
For a more in-depth look and a case-study in the design of a lock-free data structure and how it can be used in practice,
we would like to refer the reader to our chapter in ``GPU Computing Gems'' \cite{LBGPUGem}, which describes
in detail how to implement a lock-free work-stealing deque and the reasoning behind the design decisions.

\subsection{Producer-Consumer Collections}

A common approach to parallelizing applications is to divide the problem into separate threads that act as either producers or consumers. The problem of synchronizing these threads and streaming of data items between them, can be alleviated by utilizing a shared collection data structure.

\subsubsection*{Bag}
\index{concurrent data structures!bag}
\index{bag}
The Bag abstract data type is a collection of items in which items can be stored and retrieved in any order. Basic operations are \op{Add} (add an item) and \op{TryRemoveAny} (remove an arbitrary chosen item). TryRemoveAny returns the item removed. Data structures with similar semantics are also called {\em buffer}, unordered collection, unordered queue, pool, and pile in the literature.

All lock-free stacks, queues and deques implicitly implements the selected bag semantics. Afek et al. \cite{AfeKNS10} presented an explicit pool data structure. It is lock-free, although not linearizable, utilizes distributed storage and is based on randomization to establish a probabilistic level of disjoint-access-parallelism.

\index{flat-sets}
In \cite{GidPT05, GIdPT10} a data structure called {\em flat-sets}, was introduced
and used as a building block in the concurrent memory allocation service. This
is a bag-like structure that supports lock-free insertion and removal of
items as well as an ``inter-object'' operation, for moving an item from one
flat-set to another in a lock-free and linearizable manner, thus offering the
possibility of combining data structures.

In \cite{SundGPT11} a lock-free bag implementation is presented;
the algorithm
supports multiple producers and multiple consumers, as well as dynamic collection sizes. To handle concurrency
efficiently, the algorithm was designed to optimize for disjoint-access-parallelism for the supported semantics.

\subsubsection*{Stack}
\index{concurrent data structures!stack}
\index{stack}
The Stack abstract data type is a collection of items in which only the most recently added item may be removed. The latest added item is at the top. Basic operations are \op{Push} (add to the top) and \op{Pop} (remove from the top). Pop returns the item removed. The data structure is also known as a ``last-in, first-out'' or LIFO buffer.

Treiber presented a lock-free stack (a.k.a. IBM Freelist) based on linked lists, which was later efficiently fixed from the ABA problem by Michael \cite{Mic04b}. Also Valois \cite{Val95} presented a lock-free implementation that uses the \op{CAS} atomic primitive. Hendler et al. \cite{HenSY10} presented an extension where randomization and \index{elimination} elimination are used for increasing scalability when \index{contention} contention is detected on the \op{CAS} attempts.

\subsubsection*{Queue}
\index{concurrent data structures!queue}
\index{queue}
The Queue abstract data type is a collection of items in which only the earliest added item may be accessed. Basic operations are \op{Enqueue} (add to the tail) and \op{Dequeue} (remove from the head). Dequeue returns the item removed. The data structure is also known as a ``first-in, first-out'' or FIFO buffer.

\index{wait-free}
Lamport \cite{Lam83} presented a lock-free (actually wait-free) implementation of a queue based on a static array, with a limited concurrency supporting only one producer and one consumer. Giacomoni et al. \cite{GiaMV08} presented a cache-aware modification which, instead of using shared head and tail indices, synchronize directly on the array elements. Herman and Damian-Iordache \cite{HerD97} outlined a wait-free implementation of a shared queue for any number of threads, although non-practical due to its high time complexity and limited capacity.

Gong and Wing \cite{GonW90} and later Shann et al. \cite{ShaHC00} presented a lock-free shared queue based on a cyclic array and the \op{CAS} primitive, though with the drawback of using version counters, thus requiring double-width \op{CAS} for storing actual items. Tsigas and Zhang \cite{TsiZ01b} presented a lock-free extension of \cite{Lam83} for any number of threads where synchronization is done both on the array elements and the shared head and tail indices using \op{CAS}, and the ABA problem is avoided by exploiting two (or more) null values.

Valois \cite{Val94,Val95} makes use of linked lists in his lock-free implementation which is based on the \op{CAS} primitive. Prakash et al. \cite{PraLJ94} also presented an implementation using linked lists and the \op{CAS} primitive, although with the drawback of using version counters and having low scalability. Michael and Scott \cite{MicS96} presented a lock-free queue that is more efficient, synchronizing via the shared head and tail pointers as well via the next pointer of the last node. Moir et al. \cite{MoiNSS05} presented an extension where elimination is used as a \index{backoff} back-off strategy when contention on \op{CAS} is noticed, although elimination is only possible when the queue contains very few items. Hoffman et al. \cite{HofSS07} takes another approach for a back-off strategy by allowing concurrent \op{Enqueue} operations to insert the new node at adjacent positions in the linked list if contention is noticed.
Gidenstam et al. \cite{GidST10} combines the efficiency of using arrays and the dynamic capacity of using linked lists, by providing a lock-free queue based on linked lists of arrays, all updated using \op{CAS} in a cache-aware manner.

\subsubsection*{Deque}
\index{concurrent data structures!deque}
\index{deque}
The Deque (or doubly-ended queue) abstract data type is a combination of the stack and the queue abstract data types. The data structure is a collection of items in which the earliest as well as the latest added item may be accessed. Basic operations are \op{PushLeft} (add to the head), \op{PopLeft} (remove from the head), \op{PushRight} (add to the tail), and \op{PopRight} (remove from the tail). PopLeft and PopRight return the item removed.

\index{work-stealing}
Large efforts have been put on the work on so called work-stealing deques. These data structures only support three operations and with a limited level of concurrency, and are specifically aimed for scheduling purposes. Arora et al. \cite{AroBP98} presented a lock-free work-stealing deque implementation based on the \op{CAS} atomic primitive. Hendler et al. \cite{HenLMS06} improved this algorithm to also handle dynamic sizes.

Several lock-free implementations of the deque abstract data type for general purposes, although based on the non-available \op{CAS2} atomic primitive, have been published in the literature \cite{Gre99,ADFGMSS00,DFGMSS00,MarMS02,AttH06}. Michael \cite{Mic03} presented a lock-free deque implementation based on the \op{CAS} primitive, although not supporting any level of disjoint-access-parallelism. Sundell and Tsigas \cite{SunT08} presented a lock-free implementation that allows both disjoint-access-parallelism as well as dynamic sizes using the standard \op{CAS} atomic primitive.

\subsubsection*{Priority Queue}
\index{concurrent data structures!priority queue}
\index{priority queue}
The Priority Queue abstract data type is a collection of items which can efficiently support finding the item with the highest priority. Basic operations are \op{Insert} (add an item), \op{FindMin} (finds the item with minimum (or maximum) priority), and \op{DeleteMin} (removes the item with minimum (or maximum) priority). DeleteMin returns the item removed.

\index{wait-free}
Israeli and Rappoport \cite{IsrR93} have presented a wait-free algorithm for a shared priority queue, that requires the non-available multi-word \op{LL/SC} atomic primitives. Greenwald \cite{Gre99} has presented an outline for a lock-free priority queue based on the non-available \op{CAS2} atomic primitive. Barnes \cite{Bar92} presented an incomplete attempt for a lock-free implementation that uses atomic primitives available on contemporary systems. Sundell and Tsigas \cite{SunT05} presented the first lock-free implementation of a priority queue based on skip lists and the \op{CAS} atomic primitive.

\subsection{Lists}
\index{concurrent data structures!list}
\index{list}
The List abstract data type is a collection of items where two items are related only with respect to their relative position to each other. The data structure should efficiently support traversals among the items. Depending on what type of the underlying data structure, e.g. \textit{arrays} or \textit{linked lists}, different strengths of traversal functionality are supported.

\subsubsection*{Array}
\index{concurrent data structures!array}
\index{array}
List implementations based on the fundamental array data structure can support traversals to absolute index positions. Higher level abstractions as extendable arrays are in addition supporting stack semantics. Consequently, the Array abstract data type would support the operations \op{ReadAt} (read the element at index), \op{WriteAt} (write the element at index), \op{Push} (add to the top) and \op{Pop} (remove from the top). Pop returns the item removed.

A lock-free extendable array for practical purposes has been presented by
Dechev et al. \cite{DecPS06}.

\subsubsection*{Linked List}
\index{concurrent data structures!linked-list}
\index{linked-list}
In a concurrent environment with List implementations based on linked lists, traversals to absolute index positions are not feasible. Consequently, traversals are only supported relatively to a current position. The current position is maintained by the cursor concept, where each handle (i.e. thread or process) maintains one independent cursor position. The first and last cursor positions do not refer to real items, but are instead used as end markers, i.e. before the first item or after the last item. Basic operations are \op{InsertAfter} (add a new item after the current), \op{Delete} (remove the current item), \op{Read} (inspect the current item), \op{Next} (traverse to the item after the current), \op{First} (traverse to the position before the first item). Additional operations are \op{InsertBefore} (add a new item before the current), \op{Previous} (traverse to the item before the current), and \op{Last} (traverse to the position after the last item).

Lock-free implementations of the singly-linked list based on the \op{CAS} atomic primitive and with semantics suitable for the Dictionary abstract type rather than the List has been presented by Harris \cite{Har01}, Michael \cite{Mic02}, and Fomitchev and Ruppert \cite{FomR04}. Greenwald \cite{Gre02} presented a doubly-linked list implementation of a dictionary based on the non-available \op{CAS2} atomic primitive. Attiya and Hillel \cite{AttH06} presented a \op{CAS2}-based implementation that also supports disjoint-access-parallelism. Valois \cite{Val95} outlined a lock-free doubly-linked list implementation with all list semantics except delete operations. A more general doubly-linked list implementation supporting general list semantics was presented by Sundell and Tsigas \cite{SunT08}.

\subsection{Sets and Dictionaries}
\index{concurrent data structures!set}\index{concurrent data structures!dictionary}
\index{set} \index{dictionary}
The Set abstract data type is a collection of special items called \textit{keys}, where each key is unique and can have at most one occurrence in the set.
 Basic operations are \op{Add} (adds the key), \op{ElementOf} (checks if key is present), and \op{Delete} (removes the key).

The Dictionary abstract data type is a collection of items where each item is associated with a unique key. The data structure should efficiently support finding the item associated with the specific key. Basic operations are \op{Insert} (add an item associated with a key), \op{Find} (finds the item associated with a certain key), and \op{Delete} (removes the item associated with a certain key). Delete returns the item removed. In concurrent environments, an additional basic operation is the \op{Update} (re-assign the association of a key with another item) operation.

Implementations of Sets and Dictionaries are often closely related in a way that most implementations of a set can be extended to also support dictionary semantics in a straight forward manner. However, the \op{Update} operation mostly needs specific care in the fundamental part of the algorithmic design to be linearizable. Non-blocking implementations of sets and dictionaries are mostly based on hash-tables or linked lists as done by Valois \cite{Val95}. The path using concurrent linked lists was improved by Harris \cite{Har01}. Other means to implement sets and dictionaries are the skip-list and tree data structures.

\subsubsection*{Skip-List}
\index{concurrent data structures!skip-list}
\index{skip-list}
Valois \cite{Val95} outlined an incomplete idea of how to design a concurrent skip list. Sundell and Tsigas presented a lock-free implementation of a skip list in the scope of priority queues \cite{SunT03,SunT05} as well as dictionaries \cite{SunT04,Sun04} using the \op{CAS} primitive. Similar constructions have appeared in the literature by Fraser \cite{Fra03}, and Fomitchev and Ruppert \cite{FomR04}.

\subsubsection*{Hash-Table}
\index{concurrent data structures!hash-table}
\index{hash-table}
Michael \cite{Mic02} presented a lock-free implementation of the set abstract data type based on a hash-table with its chaining handled by an improved linked list compared to \cite{Har01}. To a large part, its high efficiency is thanks to the memory management scheme applied. The algorithm was improved by Shalev and Shavit \cite{ShaS03} to also handle dynamic sizes of the hash-table's underlying array data structure. Greenwald \cite{Gre02} have presented a dictionary implementation based on chained hash-tables and the non-available \op{CAS2} atomic primitive.

Gao et al. \cite{GaoGH05} presented a lock-free implementation of the dictionary abstract data type based on a hash-table data structure using open addressing. The hash-table is fully dynamic in size, although its efficiency is limited by its relatively complex memory management.

\subsubsection*{Tree}
\index{concurrent data structures!tree}
\index{tree}
Tsay and Li \cite{TsaL94} presents an approach for designing lock-free implementations of a tree data structure using the \op{LL/SC} atomic primitives and extensive copying of data. However, the algorithm is not provided with sufficient evidence for showing linearizability. Ellen et al. \cite{EllFRB10} presented a lock-free implementation of the set abstract data type based on a binary tree data structure using the \op{CAS} atomic primitive.
Spiegel and Reynolds \cite{SpiR10} presents a lock-free implementation of the set abstract data type based on a skip-tree and the \op{CAS} atomic primitive.

\section{Memory Management for Concurrent Data-Structures}
\index{memory!management}
The problem of managing dynamically allocated memory in a concurrent
environment has two parts, keeping track of the free memory available for
allocation and safely reclaim allocated memory when it is no longer in use,
i.e. memory allocation and memory reclamation.

\subsection{Memory Allocation}%
\label{subsec:memalloc}
\index{memory!allocation}
\index{malloc}
A memory allocator manages a pool of memory (heap), e.g. a contiguous
range of addresses or a set of such ranges, keeping track of which
parts of that memory are currently given to the application and which
parts are unused and can be used to meet future allocation requests
from the application. A traditional (such as the ``libc'' malloc)
general purpose memory allocator is not allowed to move or otherwise
disturb memory blocks that are currently owned by the application.

Some of the most important properties that distinguish memory allocators
for concurrent applications in the literature are:

\index{memory!fragmentation}
\noindent{\bf Fragmentation} To minimize fragmentation is to minimize
  the amount of free memory that cannot be used (allocated) by the
  application due to the size of the memory blocks.

\index{false-sharing}
\noindent{\bf False-sharing} False sharing is when different parts of
  the same cache-line are allocated to separate objects that end up
  being used by threads running on different processors.

\noindent{\bf Efficiency and scalability} The concurrent memory
  allocator should be as fast as a good sequential one when executed
  on a single processor and its performance should scale with the load
  in the system.

Here we focus on lock-free memory allocators but there is also
a considerable number of lock-based concurrent memory allocators in the
literature.


Early work on lock-free memory allocation is the work on non-blocking
operating systems by Massalin and Pu~\cite{MasP91,Mas92} and Greenwald and
Cheriton~\cite{GreC96,Gre99}.
\index{LFMalloc} \index{NBmalloc} \index{Streamflow}
Dice and Garthwaite~\cite{DiG02} presented LFMalloc, a memory
allocator based on the architecture of the Hoard lock-based concurrent
memory allocator~\cite{BeMB00} but with reduced use of locks.
Michael~\cite{Mic04} presented a fully lock-free allocator, also
loosely based on the Hoard architecture.
Gidenstam et al.~\cite{GidPT05} presented NBmalloc, another lock-free
memory allocator loosely based on the Hoard architecture. NBmalloc is
designed from the requirement that the first-remove-then-insert
approach to moving references to large internal blocks of memory
(superblocks) around should be avoided and therefore introduces and
uses a move operation that can move a reference between different
internal data-structures atomically.
Schneider et al.~\cite{SchAN06} presented Streamflow, a lock-free
memory allocator that has improved performance over previous solutions
due to allowing thread local allocations and deallocations without
synchronization.

\subsection{Memory Reclamation}%
\label{subsec:memrec}
\index{memory!reclamation}
To manage dynamically allocated memory in non-blocking algorithms is
difficult due to overlapping operations that might
read, change or \emph{dereference} (i.e. follow) references to
dynamically allocated blocks of memory concurrently.
One of the most problematic cases is when a slow
process dereferences a pointer value that it previously read from a shared
variable. This dereference of the pointer value could
occur an arbitrarily long time after the shared pointer holding that value
was overwritten and the memory designated by the pointer removed from
the shared data structure. Consequently it is impossible to safely free
or reuse the block of memory designated by this pointer value until we are
sure that there are no such slow processes with pointers to that
block.

There are several reclamation schemes in the literature with a wide and varying
range of properties:

\noindent{\bf I. Safety of local references}
  For local references, which are stored in private variables
  accessible only by one thread, to be safe the memory reclamation
  scheme must guarantee that a dynamically allocated node is never
  reclaimed while there still are local references pointing to it. 

\noindent{\bf II. Safety of shared references}
  Additionally, a memory reclamation scheme could also guarantee that
  it is always safe for a thread to dereference any shared references
  located within a dynamic node the thread has a local reference to.
  Property~I alone does not guarantee this, since for a node that has
  been deleted but cannot be reclaimed yet any shared references
  within it could reference nodes that have been deleted and reclaimed
  since the node was removed from the data structure.

\noindent{\bf III. Automatic or explicit deletion}
  A dynamically allocated node could either be reclaimed automatically
  when it is no longer accessible through any local or shared
  reference, that is, the scheme provides \emph{automatic garbage
    collection}, or the user algorithm or data structure could be
  required to explicitly tell the memory reclamation scheme when a
  node is removed from the active data structure and should be
  reclaimed as soon as it has become safe. While automatic garbage
  collection is convenient for the user, explicit deletion by the user
  gives the reclamation scheme more information to work with and can help to
  provide stronger guarantees, e.g. bounds on the amount of deleted
  but yet unreclaimed memory.

\noindent{\bf IV. Requirements on the memory allocator}
  Some memory reclamation schemes require special properties from the
  memory allocator, like, for example, that each
  allocable node has a permanent (i.e. for the rest of the system's
  lifetime) reference counter associated with it. Other schemes are
  compatible with the well-known and simple \op{allocate}/\op{free}
  allocator interface where the node has ceased to exist after the
  call to \op{free}.

\noindent{\bf V. Required synchronization primitives}
  Some memory reclamation schemes are defined using synchronization
  primitives that few if any current processor architectures provide
  in hardware, such as for example \emph{double word}
  \op{CAS}, which then have to be implemented in
  software often adding considerable overhead. Other schemes make do
  with \emph{single word} \op{CAS}, \emph{single word}
  \op{LL/SC} or even just reads and
  writes alone.

\begin{table}
  \renewcommand{\thefootnote}{\alph{footnote}}
    \begin{small}
    \begin{tabular}{|l|c|c|c|c|}
      \hline
      & Property~II
      & Property~III
      & Property~IV
      & Property~V\\
      \hline
      Michael \cite{Mic02b,Mic04b}
      &  No & Explicit  & Yes & Yes\\ 
      Herlihy et al. \cite{HerLM02}
      &  No & Explicit  & Yes & No\\ 
      Valois et al. \cite{Val95,MicS95}
      & Yes & Automatic &  No & Yes\\ 
      Detlefs et al. \cite{DetMMS01}
      & Yes & Automatic & Yes & No\\ 
      Herlihy et al. \cite{HerLMM:2005:NMM}
      & Yes & Automatic & Yes & No\\ 
      Gidenstam et al. \cite{GidPST05,GidPST09}
      & Yes & Explicit & Yes & Yes\\ 
      Fraser \cite{Fra03}
      & Yes & Explicit  & Yes & Yes\\ 
      Herlihy et al. \cite{HerM92a}
      & Yes & Automatic & Integrated & Yes\\ 
      Gao et al. \cite{GaoGH07}
      & Yes & Automatic & Integrated & Yes\\ 
      \hline
    \end{tabular}
  \end{small}
  \caption{Properties of different approaches to non-blocking
    memory reclamation.}
  \label{table:lfmr-comparison}
\end{table}
The properties of the memory reclamation schemes discussed here are
summarized in Table~\ref{table:lfmr-comparison}.
One of the most important is Property~II,
which many lock-free
algorithms and data structures need.  Among the memory
reclamation schemes that guarantee Property~II we have the following
ones, all based on reference counting:
Valois et al.~\cite{Val95,MicS95},
Detlefs et al.~\cite{DetMMS01}, Herlihy et al.~\cite{HerLMM:2005:NMM}
and Gidenstam et al.~\cite{GidPST05,GidPST09} and
the potentially blocking epoch-based scheme by Fraser~\cite{Fra03}.

On the other hand, for data structures that do not need Property~II,
for example stacks, the use of a reclamation scheme that does not
provide this property has significant potential to offer reduced overhead
compared to the stronger schemes.
Among these memory reclamation schemes we have the non-blocking ones
by Michael~\cite{Mic02b,Mic04b} and
Herlihy et al.~\cite{HerLM02}.

\paragraph{\bf Fully Automatic Garbage Collection}
\index{garbage collection}
A fully automatic garbage collector provides property I, II and III
with automatic deletion.

There are some lock-free garbage collectors in the literature. Herlihy
and Moss presented a lock-free copying garbage collector in
\cite{HerM92a}.  Gao et al.~\cite{GaoGH07} presented a lock-free Mark
\& Sweep garbage collector and Kliot at al.~\cite{KliPS09} presented a
lock-free stack scanning mechanism for concurrent garbage collectors.

\section{Graphics Processors}\label{SecGpuorComposing}
\index{GPU} \index{CUDA} \index{OpenCL}
Currently the two most popular programming environments for general purpose computing for graphics processors
are CUDA and OpenCL. 
Neither
provides any direct support for locks, and it is unlikely that this will change in the future. 
Concurrent data structures that are used on graphics processors will therefore have to be lock-free.

While graphics processors share many features with conventional processors, and many lock-free algorithms can be
ported directly, there are some differences that are important to consider, if one also wants to maintain or
improve the scalability and throughput of the algorithms.

\subsection{Data Parallel Model}
\index{SIMD computing}
A graphics processor consists of a number of multiprocessors that can execute the same instruction
on multiple data, known as SIMD computing. 
Concurrent data structures are, as the name implies, designed to support
multiple concurrent operations, but when used on a multiprocessor they also need to support concurrent
instructions within an operation. This is not straightforward, as most have been designed for scalar processors.
Considering that SIMD instructions play an instrumental role in the parallel performance offered by the
graphics processor, it is imperative that this issue be addressed.

Graphics processor have a wide memory bus and a high memory bandwidth, which makes it possible to quickly transfer
data from the memory to the processor and back. The hardware is also capable of coalescing multiple small
memory operations into a single, large, atomic memory operation.
As a single large memory operation can be performed faster than many small, this should be
taken advantage of in the algorithmic design of the data structure.

The cache in graphics processors is smaller than on conventional SMP processors and in many cases non-existent. The memory latency is
instead masked by utilizing thousands of threads and by storing data temporally in a high-speed multiprocessor
local memory area.
The high number of threads reinforces
the importance of the data structure being highly scalable.

\index{livelock} \index{hardware scheduling}
The scheduling of threads on a graphics processor is commonly being performed by the hardware.
Unfortunately, the scheme used is often undocumented, thus there is no guarantee that it will be fair.
This makes the use of algorithms with blocking behavior risky.
For example, a thread holding a lock could be indefinitely swapped out in favor of another thread waiting for the same lock,
resulting in a livelock situation. Lock-freeness is thus a must.

Of a more practical concern is the fact that a graphics processor often lacks stacks, making recursive operations
more difficult. The lack of a joint address space between the GPU and the CPU also complicates the move of data
from the CPU to the graphics processor, as all pointers in the data structure have to be rebased when moved to a new address.

\subsection{New Algorithmic Design}
 The use of SIMD instructions means that if multiple threads write to the same memory location,
 only one (arbitrary) thread can succeed. Thus, allowing threads that will be combined to a SIMD unit
 by the hardware to concurrently try to enqueue an item to the same position in a queue, will
with all likelihood be unnecessarily expensive,
as only one thread can succeed in enqueing its item. Instead, by first combining the operations locally, and then trying
to insert all elements in one step, this problem can be avoided. This is a technique used by XMalloc, a lock-free memory
allocator for graphics processors \cite{xmalloc}. On data structures with more disjoint memory access than a queue, the
problem is less pronounced, as multiple operations can succeed concurrently if they access different parts of the memory.

 An example of a way to take advantage of the SIMD instructions and memory coalescing, is to allow each node
in a tree to have more children. Allowing a node in a tree to have more children will have the effect of making the
tree shallower and lower the number of nodes that needs to checked when searching for an item. As a consequence,
the time spent in each node will increase, but with coalesced memory access and SIMD instructions, this increase
in time spent can be limited by selecting the number of children to suit the SIMD instruction size. The node can
then be read in a single memory operation and the correct child can be found using just two SIMD compare instructions.

\index{lazy operations}
 Another suggestion is to use memory coalescing to implement lazy operations, where larger read and write operations
replace a percentage of expensive CAS operations. An array-based queue for example does not need to update
its tail pointer using CAS every time an item is inserted. Instead it could be updated every $x$:th operation, and the
correct tail could be found by quickly traversing the array using large memory reads and SIMD instructions,
reducing the traversal time to a low static cost. This type of lazy updating was used in the queue by
Tsigas and Zhang \cite{TsiZ01}.

\index{coalescing} 
 The coalescing memory access mechanism also directly influences the synchronization capabilities of the graphics processor. It has for example been shown that it can be used to facilitate wait-free synchronization between threads, without the need of synchronization primitives other than reads and writes \cite{HaTA08_IPDPS, HaTA_TPDS09}.

\index{load balancing} \index{work-stealing}
When it comes to software-controlled load balancing, there have been experiments made comparing the built-in hardware scheduler with a software managed work-stealing approach \cite{loadbalgpu}. It was shown that lock-free implementations of data structures worked better than lock-based, and that lock-free work-stealing could outperform
the built-in scheduler.


The lack of a stack can be a significant problem for data structures that require recursive helping for
lock-freeness. While it is often possible to rewrite recursive code to work iteratively instead, it requires
that recursive depth can be bounded to lower the amount of memory that needs to be allocated.



\bibliographystyle{plain}
\bibliography{BookChapter}      

\end{document}